\documentclass[twocolumn,aps,pra,amsmath,amssymb,nofootinbib,showpacs,superscriptaddress]{revtex4-1}
\usepackage[latin9]{inputenc}
\usepackage{amstext}
\usepackage{amssymb}
\usepackage{graphicx}
\usepackage{esint}

\makeatletter
\usepackage{epsfig}
\usepackage{bm}
\usepackage{dcolumn}
\usepackage{color}
\usepackage{cleveref}

\makeatother

\begin{document}

\title{Attractive Fermi polarons at nonzero temperature with finite impurity
concentration}

\author{Hui Hu}

\affiliation{Centre for Quantum and Optical Science, Swinburne University of Technology,
Melbourne, Victoria 3122, Australia}

\author{Brendan C. Mulkerin}

\affiliation{Centre for Quantum and Optical Science, Swinburne University of Technology,
Melbourne, Victoria 3122, Australia}

\author{Jia Wang}

\affiliation{Centre for Quantum and Optical Science, Swinburne University of Technology,
Melbourne, Victoria 3122, Australia}

\author{Xia-Ji Liu}

\affiliation{Centre for Quantum and Optical Science, Swinburne University of Technology,
Melbourne, Victoria 3122, Australia}

\date{\today}
\begin{abstract}
We theoretically investigate how quasi-particle properties of an attractive
Fermi polaron are affected by nonzero temperature and finite impurity
concentration. By applying both non-self-consistent and self-consistent
many-body $T$-matrix theories, we calculate the polaron energy (including
decay rate), effective mass, and residue, as functions of temperature
and impurity concentration. The temperature and concentration dependences
are weak on the BCS side with a negative impurity-medium scattering
length. Toward the strong attraction regime across the unitary limit,
we find sizable dependences. In particular, with increasing temperature
the effective mass quickly approaches the bare mass and the residue
is significantly enhanced. At the temperature $T\sim0.1T_{F}$, where
$T_{F}$ is the Fermi temperature of the background Fermi sea, the
residual polaron-polaron interaction seems to become attractive. This
leads to a notable down-shift in the polaron energy. We show that,
by taking into account the temperature and impurity concentration
effects, the measured polaron energy in the first Fermi polaron experiment
{[}A. Schirotzek \textit{et al.}, Phys. Rev. Lett. \textbf{102}, 230402
(2009){]} can be better theoretically explained. 
\end{abstract}

\maketitle

\section{Introduction}

Over the past two decades ultracold atomic gases have provided an
ideal platform to understand the intriguing behavior of quantum many-body
systems \cite{Bloch2008}. The simplest example of an interacting
many-body system is a moving impurity immersed in a background medium
\cite{Massignan2014}. In this so-called polaron problem, the interaction
between impurity and medium in an ultracold Fermi gas can be tuned
arbitrarily by using Feshbach resonances \cite{Chin2010}. The motion
of the impurity is then addressed by low-energy excitations of the
background medium and its fundamental properties are profoundly affected
\cite{Massignan2014}. In the case of fermionic impurity with finite
density/concentration, the emergence of Fermi liquid behavior is anticipated
\cite{Lobo2006,Pilati2008}.

Theoretically, it turns out that the polaron problem can be well approximated
by a variational ansatz that includes only one-particle-hole excitations,
as proposed by Chevy in 2006 in his seminal work \cite{Chevy2006}.
For a non-interacting single-component Fermi sea as the background
medium, and when the impurity-medium scattering length $a$ is tuned
to the unitary limit ($a\rightarrow\infty$), Chevy's ansatz predicts
a polaron energy of $E_{P}\simeq-0.607\varepsilon_{F}$ \cite{Chevy2006,Combescot2007},
where $\varepsilon_{F}$ is the Fermi energy of the Fermi sea, which
is comparable to the numerically exact diagrammatic Monte Carlo (Diag-MC)
result of $E_{P}=-0.615(1)\varepsilon_{F}$ \cite{Prokofev2008,Vlietinck2013,Kroiss2015}.
This excellent agreement shows the physically important contributions
to the polaron energy and may result from a cancellation of the higher-order
contributions, as was investigated by the next order calculation with
the inclusion of two-particle-hole excitations \cite{Combescot2008}.
The simple variational ansatz was later used to discover repulsive
Fermi polarons in a meta-stable upper branch \cite{Cui2010} and to
describe Bose polarons \cite{Li2014}. At the level of including two-particle-hole
excitations, Chevy's ansatz has also been applied to systems with
bosonic degrees of freedom, such as a Bose-Einstein condensate (BEC)
\cite{Levinsen2015} or a Bardeen\textendash Cooper\textendash Schrieffer
(BCS) superfluid \cite{Nishida2015,Yi2015}. In those cases, the interplay
between polarons and the resulting Efimov trimer was explored \cite{Levinsen2015,Nishida2015,Yi2015,Sun2017}.

Experimentally, the first realization of attractive Fermi polarons
was carried out by the group at Massachusetts Institute of Technology
(MIT) in 2009 using $^{6}$Li atoms \cite{Schirotzek2009}. The polaron
energy and residue were determined by using radio-frequency (rf) spectroscopy
in the vicinity of the unitary limit and the strong attractive BEC
regime. The attractive polaron picture was used later to understand
the radio-frequency spectrum of a quasi-two-dimensional Fermi gas
\cite{Zhang2012}. The existence of metastable repulsive Fermi polarons
was experimentally confirmed in 2012 by immersing $^{40}$K impurity
in a Fermi sea of $^{6}$Li atoms near a narrow Feshbach resonance
\cite{Kohstall2012}, and in two dimensions by using $^{40}$K atoms
\cite{Koschorreck2012}. Most recently, a careful analysis of the
quasi-particle properties of repulsive Fermi polarons in $^{6}$Li
systems was performed at the European Laboratory for Non-linear Spectroscopy
(LENS), Florence \cite{Scazza2017}. The experimental observation
of attractive and repulsive Bose polarons has also been reported \cite{Hu2016,Jorgensen2016}.

In these experiments the data was compared with the theoretical predictions
of a single impurity at zero temperature \cite{Massignan2014}. The
unavoidable nonzero temperature and finite impurity concentration
in real experiments are anticipated to give negligible corrections.
However, those corrections have never been carefully examined, except
the idealized case of 1D Fermi polarons \cite{Doggen2013}, where
the exact solution is available. A possible reason is that the current
polaron theory relies heavily on Chevy's variational approach \cite{Chevy2006},
which unfortunately is difficult to extend to nonzero temperature
and finite impurity concentration \cite{Parish2016}.

The purpose of the present work is to address the effects of nonzero
temperature and finite impurity concentration for attractive Fermi
polarons, by using both non-self-consistent \cite{Combescot2007,Massignan2008,Punk2009,Massignan2011,Schmidt2012,Baarsma2012}
and self-consistent $T$-matrix theories \cite{Haussmann1994,Liu2005,Haussmann2007,Rath2013}.
In the limit of zero temperature and a single impurity, the non-self-consistent
$T$-matrix theory is equivalent to Chevy's variational approach \cite{Combescot2007,Schmidt2012}.
At weak attractions on the BCS side of the impurity and medium scattering
resonance, we find that the effects of nonzero temperature and finite
impurity concentration are indeed negligible, confirming the previous
anticipation. However, across the resonance and toward the strong
attraction regime, a nonzero temperature may significantly reduce
the effective mass and enhance the residue of attractive Fermi polarons.
The polaron energy also shows a considerable temperature dependence.
In particular, at the typical experimental temperatures $T\sim0.1T_{F}$,
where $T_{F}$ is the Fermi temperature of the Fermi sea, the polaron-polaron
interaction may become attractive, leading to a sizable downshift
in the polaron energy, which may be experimentally resolved. Indeed,
by taking into account the temperature and impurity concentration
effects in our non-self consistent $T$-matrix theory we find that
the measured polaron energy in the first Fermi polaron experiment
at MIT \cite{Schirotzek2009} could be better theoretically understood.

The paper is set out as follows. In Sect.~\ref{sec:Theory} we outline
the $T$-matrix theories and methodology, defining the quasi-particle
properties of Fermi polarons. In Sect.~\ref{sec:Zero_T} we give
a brief review of the zero temperature behavior of the Fermi polaron.
In Sect.~\ref{sec:Finite_T} we describe the quasi-particle properties
of the Fermi polaron at finite temperature. In Sect.~\ref{sec:Density}
we consider finite impurity density and the Fermi liquid behavior,
comparing to experimental results at finite temperature and impurity.
Finally, in Sect.~\ref{sec:conc} we summarize our results.

\section{Many-body $T$-matrix theories of attractive Fermi polarons}

\label{sec:Theory}

We consider a two-component Fermi gas of mass $m$ with a large spin
polarization (i.e., $n_{\uparrow}=n\gg n_{\downarrow}$), which is
described by the model single channel Hamiltonian \cite{Chevy2006,Combescot2007},
\begin{eqnarray}
H & = & \sum_{\mathbf{k}}\left[\left(\epsilon_{\mathbf{k}}-\mu\right)c_{\mathbf{k}\uparrow}^{\dagger}c_{\mathbf{k}\uparrow}+\left(\epsilon_{\mathbf{k}}-\mu_{\downarrow}\right)c_{\mathbf{k}\downarrow}^{\dagger}c_{\mathbf{k}\downarrow}\right]\nonumber \\
 &  & +\frac{U}{V}\sum_{\mathbf{q},\mathbf{k},\mathbf{k}'}c_{\mathbf{k}\uparrow}^{\dagger}c_{\mathbf{q}-\mathbf{k}\downarrow}^{\dagger}c_{\mathbf{q}-\mathbf{k}'\downarrow}c_{\mathbf{k}'\uparrow},\label{eq:hami}
\end{eqnarray}
where $\epsilon_{\mathbf{k}}\equiv\hbar^{2}\mathbf{k}^{2}/(2m)$,
$\mu$ and $\mu_{\downarrow}$ are the chemical potentials of spin-up
and spin-down atoms, respectively, and $U<0$ is the bare attractive
interatomic interaction strength, to be renormalized in terms of the
$s$-wave scattering length $a$, according to 
\begin{equation}
\frac{1}{U}=\frac{m}{4\pi\hbar^{2}a}-\sum_{\mathbf{k}}\frac{m}{\hbar^{2}\mathbf{k}^{2}}.\label{eq:renormalization}
\end{equation}
In the large spin polarization limit, we treat the spin-down atoms
as impurities and assume that, at the first order of the impurity
concentration $x=n_{\downarrow}/n$, the background medium of spin-up
atoms is not affected by interactions. As a result, $\mu$ can be
taken as the chemical potential of an ideal Fermi gas at low temperatures,
$\mu^{(0)}(T)\simeq\varepsilon_{F}=\hbar^{2}(6\pi^{2}n)^{2/3}/(2m)$,
and the thermal Green's function of spin-up atoms is 
\begin{equation}
G_{\uparrow}^{(0)}\left(\mathbf{k},i\omega_{m}\right)=\frac{1}{i\omega_{m}-\left(\epsilon_{\mathbf{k}}-\mu\right)},\label{eq: gf0up}
\end{equation}
with fermionic Matsubara frequencies $\omega_{m}\equiv(2m+1)\pi k_{B}T$
for integer $m$. The impurity chemical potential, $\mu_{\downarrow}<0$,
and the impurity thermal Green's function, $G_{\downarrow}(\mathbf{k},i\omega_{m})$,
strongly depend on the interatomic interaction and these effects must
be taken into account. For a \emph{single} impurity at \emph{zero}
temperature, $\mu_{\downarrow}$ gives the polaron energy $E_{P}$
\cite{Combescot2007}. 

\subsection{Many-body $T$-matrix theories}

We solve the impurity thermal Green's function 
\begin{equation}
G_{\downarrow}\left(\mathbf{k},i\omega_{m}\right)=\frac{1}{i\omega_{m}-\left(\epsilon_{\mathbf{k}}-\mu_{\downarrow}\right)-\sum\left(\mathbf{k},i\omega_{m}\right)}\label{eq: gfdown}
\end{equation}
by using the well-established many-body $T$-matrix theories \cite{Loktev2001,Hu2008,Mulkerin2015},
which amount to summing up all the ladder-type diagrams. In this approximation,
the self-energy $\Sigma$ of the impurity atom is given by \cite{Combescot2007},
\begin{equation}
\Sigma=k_{B}T\sum_{\mathbf{q},i\nu_{n}}G_{\uparrow}^{(0)}\left(\mathbf{q}-\mathbf{k},i\nu_{n}-i\omega_{m}\right)\Gamma\left(\mathbf{q},i\nu_{n}\right),\label{eq:selfenergy1}
\end{equation}
where the vertex function $\Gamma$ can be written through the Bethe-Salpeter
equation as, 
\begin{equation}
\Gamma\left(\mathbf{q},i\nu_{n}\right)=\frac{1}{U^{-1}+\chi\left(\mathbf{q},i\nu_{n}\right)},\label{eq:vertexfunction}
\end{equation}
and the pair propagator $\chi(\mathbf{q},i\nu_{n})$ is 
\begin{equation}
\chi=k_{B}T\sum_{\mathbf{k},i\omega_{m}}G_{\uparrow}^{(0)}\left(\mathbf{q}-\mathbf{k},i\nu_{n}-i\omega_{m}\right)G_{\downarrow}\left(\mathbf{k},i\omega_{m}\right).\label{Eq:propagator2p}
\end{equation}
Here, $\nu_{n}\equiv2n\pi k_{B}T$ with integer $n$ are the bosonic
Matsubara frequencies. \Cref{eq: gfdown,eq:selfenergy1,eq:vertexfunction,Eq:propagator2p}
form a closed set of equations, which have to be solved self-consistently.
We refer to this set of equations as the self-consistent $T$-matrix
theory, or the ``$G_{0\uparrow}G_{\downarrow}$'' theory, according
to the structure in the pair propagator. We note that in a strong-coupling
theory, the self-consistency does not necessarily guarantee a better
or more accurate theory \cite{Hu2008,Mulkerin2015}. For the calculation
of the energy of attractive Fermi polarons at zero temperature, it
is actually more useful to take a non-self-consistent treatment of
the many-body $T$-matrix theory, as suggested by the comparison with
numerically exact Diag-MC simulations in three dimensions \cite{Prokofev2008,Vlietinck2013,Kroiss2015,Combescot2008}
or exact Bethe ansatz solutions in one dimension \cite{Doggen2013}.
That is, we simply use a non-interacting impurity Green's function
\begin{equation}
G_{\downarrow}^{(0)}\left(\mathbf{k},i\omega_{m}\right)=\frac{1}{i\omega_{m}-\left(\epsilon_{\mathbf{k}}-\mu_{\downarrow}\right)}\label{eq: gf0down}
\end{equation}
in the pair propagator Eq. (\ref{Eq:propagator2p}). At zero temperature
and in the single-impurity limit, this non-self-consistent $T$-matrix
theory, or ``$G_{0\uparrow}G_{0\downarrow}$'' theory, exactly recovers
Chevy's variational approach \cite{Combescot2007}. 

In this work, we explore both non-self-consistent and self-consistent
$T$-matrix theories, as both of them are not justified in the strong-coupling
limit \cite{Hu2008,Mulkerin2015} and they may give complementary
information on some specific observables of interest. Of course, the
self-consistent calculations are much more numerically involved, since
the integration over ($\mathbf{q},i\nu_{n}$) in Eq. (\ref{eq:selfenergy1})
or ($\mathbf{k},i\omega_{m}$) in Eq. (\ref{Eq:propagator2p}) is
three-dimensional. To overcome this numerical integration difficulty,
we rewrite the self-energy and the pair propagator in real space and
in imaginary-time space \cite{Haussmann1994,Liu2005,Haussmann2007}:
\begin{equation}
\Sigma(\mathbf{x},\tau)=G_{\uparrow}^{(0)}(-\mathbf{x},-\tau)\Gamma(\mathbf{x},\tau),\label{eq:selfenergy2}
\end{equation}
and 
\begin{equation}
\chi(\mathbf{x},\tau)=G_{\uparrow}^{(0)}(\mathbf{x},\tau)G_{\downarrow}(\mathbf{x},\tau),\label{eq:propagator2p2}
\end{equation}
where in the last equation $G_{\downarrow}(\mathbf{x},\tau)$ should
be replaced with $G_{\downarrow}^{(0)}(\mathbf{x},\tau)$, if we consider
the non-self-consistent $T$-matrix theory. Thus, it is straightforward
to calculate self-energy or pair propagator once we know the Green's
functions and vertex function in real space. The trade-off is that
we now need to perform two Fourier transforms, $\mathbf{x}\longleftrightarrow\mathbf{k}$
(or $\mathbf{q}$) and $\tau\longleftrightarrow i\omega_{m}$ (or
$i\nu_{n}$) \cite{Haussmann1994,Liu2005,Haussmann2007}. Due to the
spatial homogeneity and rotational invariance, all the functions in
real space (or momentum space) depend on $x=\left|\mathbf{x}\right|$
(or $k=\left|\mathbf{k}\right|$) only. Hence, the two Fourier transforms
are essentially a two-dimensional integration and can be performed
very efficiently. It turns out that the only difficulty in our numerical
calculations is the singularities of Green functions and vertex function
near $\mathbf{x}=0$ and $\tau=0^{-}$. Fortunately, the same singularities
appear in the free Green's function $G^{(0)}$ or in the first-order
iterated vertex function $\Gamma^{(n=1)}$ and thus can be easily
taken into account in an analytic way \cite{Haussmann1994,Liu2005}.

We note that, at zero temperature in the single impurity limit, both
non-self-consistent and self-consistent $T$-matrix theories have
been implemented, as a by-product of the Diag-MC simulations of Fermi
polarons \cite{Vlietinck2013}. Here we extend these theories to the
case of nonzero temperature and finite impurity concentration.

\subsection{Quasi-particle properties of Fermi polarons}

\begin{figure}
\centering{}\includegraphics[width=0.48\textwidth]{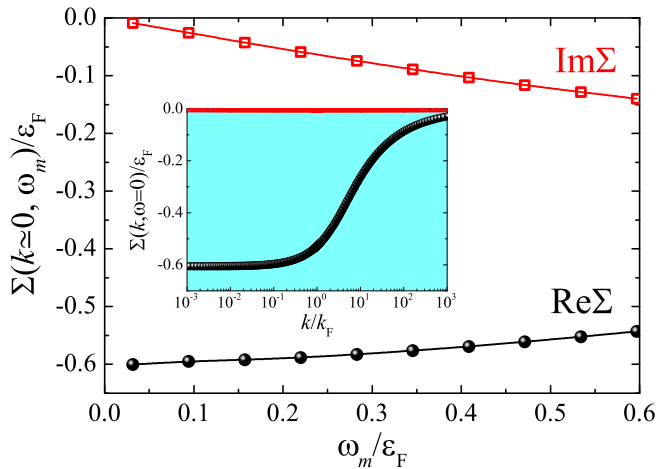}\caption{\label{fig1_selfenergy} (color online). The Matsubara frequency dependence
of the self-energy of the impurity temperature Green function at very
small momentum $k\simeq0$, calculated by using the non-self-consistent
$T$-matrix theory. Here, we take $1/(k_{F}a)=0$ (i.e., the unitary
limit), $T=0.01T_{F}$ and the impurity chemical potential $\mu_{\downarrow}=-0.607\varepsilon_{F}$.
We extrapolate the data to zero frequency to obtain the retarded self-energy
$\Sigma^{R}(k,i\omega_{m}\rightarrow\omega=0)$, which is shown in
the inset. The imaginary part of the retarded self-energy $\textrm{Im}\Sigma^{R}(\mathbf{k},0)$
(red symbols in the inset) is essentially zero due to low temperature.}
\end{figure}

Once the impurity thermal Green's function is known, we may directly
extract the quasi-particle properties of Fermi polarons, such as the
polaron energy $E_{P}$, residue $Z$, and the effective mass $m^{*}$,
by approximating the retarded impurity Green's function $G_{\downarrow}^{R}(\mathbf{k},\omega)\equiv G_{\downarrow}(\mathbf{k},i\omega_{m}\rightarrow\omega+i0^{+})$
in the low-energy and long-wavelength limit as \cite{Prokofev2008,Vlietinck2013},
\begin{equation}
G_{\downarrow}^{R}=\frac{\mathcal{Z}}{\omega-\hbar^{2}\mathbf{k}^{2}/(2m^{*})+\mu_{\downarrow}-E_{P}+i\gamma/2}+\cdots,\label{eq:gfRdown}
\end{equation}
where $\gamma$ is the decay rate of the polaron. In the case of a
well-defined quasi-particle (i.e., $\gamma\ll\left|E_{P}\right|$),
this gives rise to a polaron spectral function $A_{\downarrow}(\mathbf{k},\omega)=-(1/\pi)\textrm{Im}G_{\downarrow}(\mathbf{k},\omega)$
\cite{Massignan2008,Baarsma2012}, 
\begin{equation}
A_{\downarrow}\left(\mathbf{k},\omega\right)=\mathcal{Z}\delta\left(\omega+\mu_{\downarrow}-\frac{\hbar^{2}\mathbf{k}^{2}}{2m^{*}}-E_{P}\right)+\cdots,\label{eq:Adown}
\end{equation}
which can be experimentally measured by using \emph{direct} rf spectroscopy
that transfers impurity atoms to a third, non-interacting hyperfine
state \cite{Schirotzek2009} or by using \emph{inverse} rf spectroscopy
that flips initially non-interacting impurity atoms (in the third
state) into the strongly interacting polaron state \cite{Scazza2017}.
The explicit expressions for the polaron energy, residue, and effective
mass may be obtained by Taylor expanding the retarded self-energy,
$\Sigma^{R}(\mathbf{k},\omega)\equiv\Sigma(\mathbf{k},i\omega_{m}\rightarrow\omega+i0^{+})$,
near $\mathbf{k}=\mathbf{0}$ and $\omega=0$. By substituting the
expansion into the retarded impurity Green's function, we find that,
\begin{eqnarray}
E_{P} & = & \left(1-\mathcal{Z}\right)\mu_{\downarrow}+\mathcal{Z}\textrm{Re}\Sigma^{R}(0,0),\label{eq:Ep}\\
\mathcal{Z} & = & \left(1-\frac{\partial\mathbf{\textrm{Re}}\Sigma^{R}}{\partial\omega}\right)^{-1},\label{eq:Residue}\\
\frac{m}{m^{*}} & = & \left(1+\frac{\partial\mathbf{\textrm{Re}}\Sigma^{R}}{\partial\epsilon_{\mathbf{k}}}\right)\left(1-\frac{\partial\mathbf{\textrm{Re}}\Sigma^{R}}{\partial\omega}\right)^{-1},\label{eq:inverseMass}\\
\gamma & = & -2\mathcal{Z}\textrm{Im\ensuremath{\Sigma^{R}(0,0)}.}
\end{eqnarray}
As the impurity concentration is given by, 
\begin{eqnarray}
n_{\downarrow} & = & \sum_{\mathbf{k}}\int d\omega f\left(\omega\right)A_{\downarrow}\left(\mathbf{k},\omega\right),\\
 & \simeq & \sum_{\mathbf{k}}\mathcal{Z}f\left(E_{P}+\frac{\hbar^{2}\mathbf{k}^{2}}{2m^{*}}-\mu_{\downarrow}\right),
\end{eqnarray}
where $f(x)=1/(e^{x}+1)$ is the Fermi distribution, it is easy to
see that at zero temperature we must have the identity $E_{P}=\mu_{\downarrow}$
for a single impurity at a vanishingly small density $n_{\downarrow}\simeq0$
\cite{Combescot2007}. By using Eq.~(\ref{eq:Ep}), we obtain the
condition 
\begin{equation}
\mu_{\downarrow}=\textrm{Re}\Sigma^{R}\left(0,0\right),
\end{equation}
which determines the impurity chemical potential for a single impurity
at zero temperature \cite{Combescot2007,Prokofev2008,Vlietinck2013}.

\begin{figure}[t]
\centering{}\includegraphics[width=0.48\textwidth]{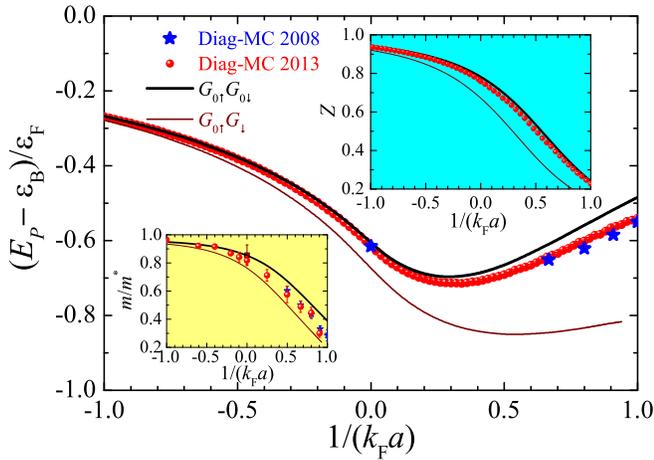}
\caption{\label{fig2_zerotemp} (color online). The energy of attractive Fermi
polarons (with the two-particle bound state energy $\epsilon_{B}=-\hbar^{2}/(ma^{2})<0$
subtracted on the BEC side), as a function of the interaction strength
$1/(k_{F}a)$, obtained by the non-self-consistent $T$-matrix theory
(thick black line) and self-consistent $T$-matix theory (thin brown
line) at $T=0.01T_{F}$, and by diagrammatic Monte Carlo simulations
at zero temperature in 2008 (blue stars) \cite{Prokofev2008} and
2013 (red circles) \cite{Vlietinck2013}. The upper and lower insets
show the residue and (inverse) effective mass of polarons, respectively.}
\end{figure}

It is worth mentioning that numerically it is difficult to directly
take the analytical continuation $i\omega_{m}\rightarrow\omega+i0^{+}$
of the retarded self-energy. At low temperatures, where $\omega_{m}\equiv(2m+1)\pi k_{B}T$
is small for small integers $m=0,1,2$, we Taylor expand the self-energy
around zero frequency, i.e., 
\begin{equation}
\Sigma\left(\mathbf{k},i\omega_{m}\right)\simeq\Sigma^{R}\left(\mathbf{k},0\right)+\frac{\partial\Sigma^{R}}{\partial\omega}\left(i\omega_{m}\right),
\end{equation}
or 
\begin{eqnarray}
\textrm{Re}\Sigma\left(\mathbf{k},i\omega_{m}\right) & \simeq & \textrm{Re}\Sigma^{R}\left(\mathbf{k},0\right)-\frac{\partial\textrm{Im}\Sigma^{R}}{\partial\omega}\omega_{m},\\
\textrm{Im}\Sigma\left(\mathbf{k},i\omega_{m}\right) & \simeq & \textrm{Im}\Sigma^{R}\left(\mathbf{k},0\right)+\frac{\partial\textrm{Re}\Sigma^{R}}{\partial\omega}\omega_{m},
\end{eqnarray}
and obtain $\Sigma^{R}(\mathbf{k},0)$ and $[\partial\Sigma^{R}(\mathbf{k},\omega)/\partial\omega]_{\omega=0}$
by numerical extrapolation. An example of such an extrapolation is
shown in Fig. \ref{fig1_selfenergy} for an attractive polaron in
the unitary limit. Consequently, we take the zero momentum limit and
calculate the numerical derivative of the retarded self-energy with
respect to $\epsilon_{\mathbf{k}}$ at $\mathbf{k}=0$.

\section{Fermi polarons at zero temperature: A brief review}

\label{sec:Zero_T}


We have calculated the polaron energy, residue, and inverse effective
mass as a function of the impurity-medium interaction strength at
nearly zero temperature, by using both non-self-consistent and self-consistent
$T$-matrix theories. These results were obtained earlier as a by-product
in a zero-temperature Diag-MC simulation \cite{Vlietinck2013}, although
the predicted polaron energy from the self-consistent $T$-matrix
theory was not explicitly reported.

\begin{figure}[t]
\centering{}\includegraphics[width=0.48\textwidth]{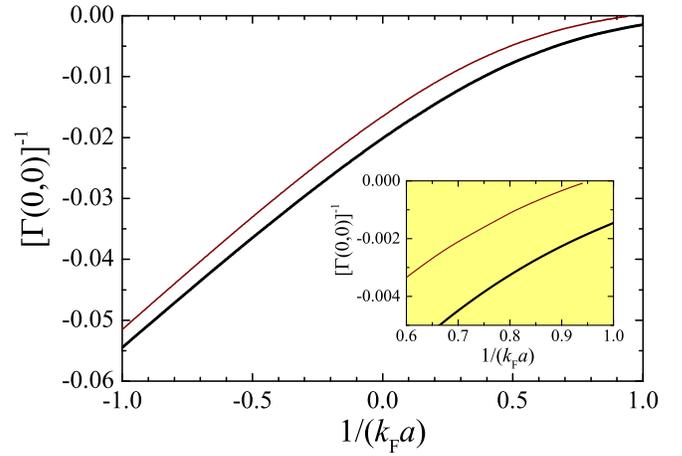}\caption{\label{fig3_thouless} (color online). The inverse of the impurity
vertex function at zero momentum and zero frequency at $T=0.01T_{F}$,
in units of $2mk_{F}$, as a function of the interaction strength
$1/(k_{F}a)$, obtained by using the non-self-consistent $T$-matrix
theory (thick black line) and self-consistent $T$-matrix theory (thin
brown line). The self-consist theory predicts a pairing instability
at the critical interaction strength $1/(k_{F}a)_{c}\simeq0.95$,
very close to the Diag-MC prediction $1/(k_{F}a)_{c}\simeq0.9$ \cite{Prokofev2008,Vlietinck2013}.
The inset is an enlarged view near the critical interaction strength.}
\end{figure}

As shown in Fig. \ref{fig2_zerotemp}, our results summarize the known
quasi-particle properties of attractive Fermi polarons at zero temperature.
For the polaron energy and residue, the prediction of the non-self-consistent
$T$-matrix theory agrees well with the numerically exact Diag-MC
simulations \cite{Prokofev2008,Vlietinck2013}. The self-consistent
$T$-matrix theory seems to underestimate the polaron energy and residue,
in particular on the BEC side, where the scattering length is positive,
$a>0$. The Diag-MC results for the inverse effective mass lies between
the predictions of the non-self-consistent and self-consistent $T$-matrix
theories, as shown in the inset at the left bottom of Fig. \ref{fig2_zerotemp}.
Overall, it is reasonable to believe that the non-self-consistent
$T$-matrix theory works better than the self-consistent theory for
attractive Fermi polarons at low temperature and small impurity concentration.
However, in some cases the self-consistent theory may provide more
useful information than the non-self-consistent theory. An interesting
example is the polaron-molecule transition, which has been predicted
by Diag-MC simulations to occur at about $1/(k_{F}a)_{c}\simeq0.9$
\cite{Prokofev2008,Vlietinck2013}.


Indeed, the self-consistent $T$-matrix theory can be used to determine
the instability of attractive Fermi polarons, with respect to the
formation of a tightly bound molecular state of an impurity atom and
a medium atom, due to their strong attraction. Numerically, we find
that beyond a threshold interaction strength $1/(k_{F}a)_{c}\simeq0.95$,
the vertex function at zero momentum and frequency becomes positive,
as shown in Fig. \ref{fig3_thouless} (thin lines). Our numerical
procedure for self-consistent calculations of the impurity Green's
function and vertex function then breaks down. Physically, it signifies
the condensation of spontaneously created molecules, following the
so-called Thouless criterion for superfluidity \cite{Nozieres1985},
\begin{equation}
\Gamma\left(\mathbf{q}=0,i\nu_{n}=0\right)=0,
\end{equation}
which is satisfied at the critical temperature $T_{c}$ or critical
interaction strength $1/(k_{F}a)_{c}$. The critical interaction strength
$1/(k_{F}a)_{c}\simeq0.95$ predicted by the self-consistent $T$-matrix
theory is comparable to the critical value obtained by Diag-MC simulations
\cite{Prokofev2008,Vlietinck2013}, in comparison, the non-self-consistent
$T$-matrix theory predicts a much larger critical interaction strength
(not shown in the figure).

\section{Fermi polarons at finite temperature}

\label{sec:Finite_T}

\begin{figure}[t]
\centering{}\includegraphics[width=0.48\textwidth]{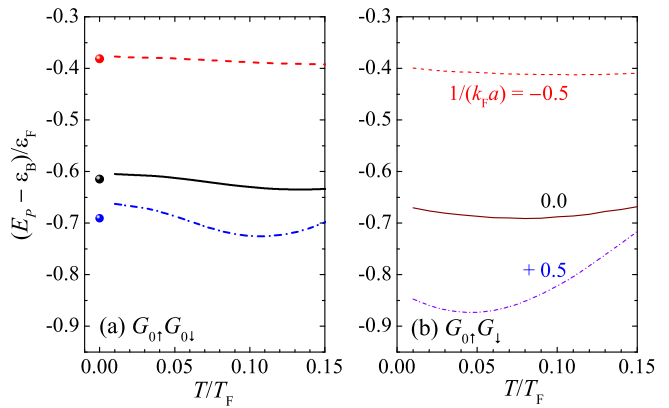}\caption{\label{fig4_ep_varyteff} (color online) Temperature dependence of
the polaron energy at the impurity concentration $x=n_{\downarrow}/n=0.01$
and at $1/(k_{F}a)=-0.5$ (dashed line), $0$ (solid line), and $0.5$
(dot-dashed line), calculated by using the non-self-consistent $T$-matrix
theory (a, left panel) and self-consistent $T$-matrix theory (b,
right panel). The circles in (a) show the diagrammatic Monte Carlo
result \cite{Vlietinck2013}.}
\end{figure}

We now turn to consider the quasi-particle properties of attractive
Fermi polarons at finite temperature. Figure \ref{fig4_ep_varyteff}
reports the polaron energy as a function of the reduced temperature
$T/T_{F}$ on the BCS side ($1/k_{F}a=-0.5$), in the unitary limit
($1/k_{F}a=0$), and on the BEC side ($1/k_{F}a=+0.5$), obtained
by using the non-self-consistent (a) and self-consistent (b) $T$-matrix
theories. At low temperatures, i.e. $T<0.05T_{F}$, the polaron energy
decreases with increasing temperature. From the viewpoint of one-particle-hole
excitations, this decrease may arise from the enlarged phase space
for particle-hole excitations at low temperature, and hence, the impurity
is dressed by more particle-hole excitations. At temperatures 
near $T\sim0.1T_{F}$ the polaron energy increases as the temperature
increases. This increase can be clearly seen on the BEC side by using
the self-consistent $T$-matrix theory, where the temperature-induced
variation of the polaron energy is about $0.1\varepsilon_{F}$ for
$T<0.15T_{F}$. Unfortunately, the current experimental measurement
of the polaron energy is not accurate enough to resolve this variation.
On the BCS side the temperature dependence of the polaron energy is
typically weak and is only about a few percent of the Fermi energy.
Both non-self-consistent and self-consistent $T$-matrix theories
predict a similar polaron energy, due to the weak attraction.

\begin{figure}[t]
\centering{}\includegraphics[width=0.48\textwidth]{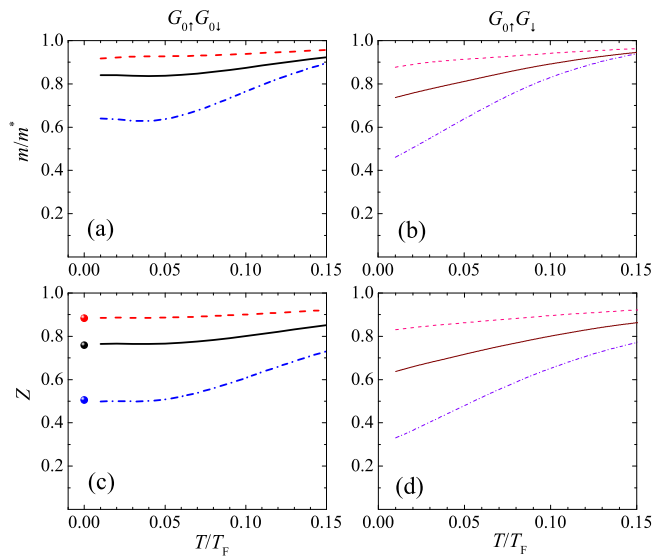}\caption{\label{fig5_MassResidue_varyteff} (color online) Temperature dependence
of the polaron effective mass (a, b) and residue (c, d) at the impurity
concentration $x=n_{\downarrow}/n=0.01$ calculated by using the non-self-consistent
$T$-matrix theory (left panel) and self-consistent $T$-matrix theory
(right panel). From top to bottom, the interaction strengths are $1/(k_{F}a)=-0.5$
(dashed line), $0$ (solid line), and $0.5$ (dot-dashed line). The
circles in (c) show the diagrammatic Monte Carlo result \cite{Vlietinck2013}.}
\end{figure}

Figure~\ref{fig5_MassResidue_varyteff} shows the temperature dependence
of the inverse effective mass (a, b) and the residue (c, d) of attractive
Fermi polarons. These two quantities increase with increasing temperature,
as predicted by both $T$-matrix theories. At $T>0.05T_{F}$, this
may be simply understood from the fact that with increasing temperature
the polaron starts to lose it polaronic character and become more
like an isolated impurity. As the temperature increases, the polarons
effective mass approaches the bare mass $m$ and its residue tends
towards unity.

\begin{figure}[t]
\centering{}\includegraphics[width=0.48\textwidth]{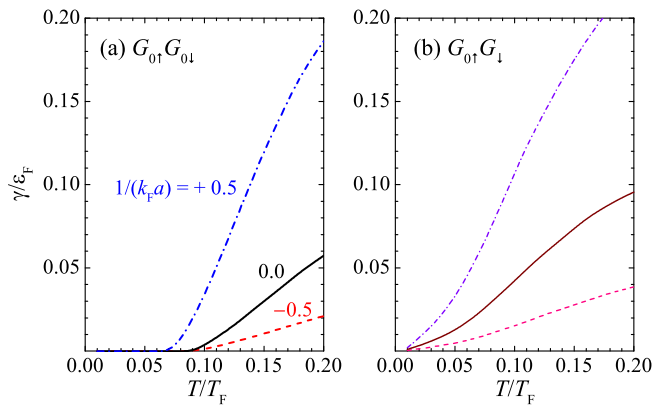}\caption{\label{fig6_decayrate_varyteff} (color online) Thermal-induced polaron
decay rate at the impurity concentration $x=n_{\downarrow}/n=0.01$
and at $1/(k_{F}a)=-0.5$ (dashed line), $0$ (solid line), and $0.5$
(dot-dashed line), calculated by using the non-self-consistent $T$-matrix
theory (a, left panel) and self-consistent $T$-matrix theory (b,
right panel).}
\end{figure}

As a well-defined quasi-particle, an attractive Fermi polaron has
infinitely long lifetime at zero temperature, unless a decay channel
to the ground-state of molecules is opened above the critical interaction
strength, $1/(k_{F}a)_{c}\sim0.9$ \cite{Bruun2010}. At finite temperature,
however, a Fermi polaron could decay via thermal excitations at arbitrary
interaction strengths, where the decay rate is anticipated to be proportional
to $(T/T_{F})^{2}$ at low temperature \cite{Schirotzek2009}. In
Fig. \ref{fig6_decayrate_varyteff}, we present the temperature dependence
of the decay rate of attractive Fermi polarons, calculated from the
non-self-consistent (a) and self-consistent (b) $T$-matrix theories.
On the BCS side, or near the unitary limit, the polaron decay rate
is less than six percent of the Fermi energy at the temperature range
considered (i.e., $T<0.2T_{F}$). In contrast, on the BEC side, the
thermal-induced decay becomes significant, where it can be as large
as $0.1\varepsilon_{F}$ at the typical experimental temperature of
$T\sim0.15T_{F}$. We note that, the decay rate determined by the
self-consistent theory is larger than the non-self-consistent theory.
Moreover, the non-self-consistent theory seems to predict a threshold
temperature (i.e., $T\sim0.08T_{F}$), below which there is no notable
decay.

\section{Fermi polarons at finite impurity density}

\label{sec:Density}

In this section, we consider the quasi-particle properties of attractive
Fermi polarons at finite impurity concentration/density and discuss
their density dependence at both essentially zero temperature (i.e.,
$T=0.01T_{F}$) and finite temperature. At the end of the section,
we also make a comparison with the first Fermi polaron experiment
\cite{Schirotzek2009}.

\begin{figure}[t]
\centering{}\includegraphics[width=0.48\textwidth]{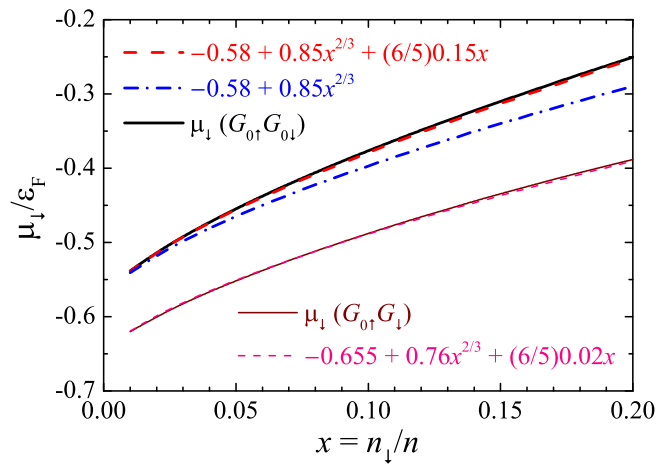}\caption{\label{fig7_FermiLiquid_varyp} (color online) The impurity chemical
potential as a function of the impurity concentration $x$ in the
unitary limit ($1/k_{F}a=0$) and at $T=0.01T_{F}$, calculated by
using the non-self-consistent $T$-matrix theory (thick black line)
and the self-consistent $T$-matrix theory (thin brown line). The
two red dashed lines are the fitting curve of the formula, $\mu_{\downarrow}=E_{P}+(m/m^{*})x^{2/3}+(6/5)\mathcal{F}x$,
where $\mathcal{F}$ is the Landau parameter characterizing the polaron-polaron
interaction. The dot-dashed line illustrates the importance of the
polaron-polaron interaction term $(6/5)\mathcal{F}x$.}
\end{figure}

\subsection{Fermi liquid behavior}

At low impurity concentration and low temperatures, Fermi polarons
are believed to form a Fermi liquid \cite{Lobo2006,Pilati2008}. In
the spirit of Landau's Fermi liquid theory, the change in the total
energy due to the addition of impurity atoms with density $x=n_{\downarrow}/n=N_{\downarrow}/N$,
where $N_{\ensuremath{\downarrow}}$ and $N$ are respectively the
number of impurity atoms and medium atoms, can be written in an energy
functional with two density dependent terms \cite{Lobo2006,Pilati2008},
\begin{equation}
\Delta E\simeq N_{\downarrow}E_{P}+\frac{3}{5}N\varepsilon_{F}\left[\frac{m}{m^{*}}\left(\frac{N_{\downarrow}}{N}\right)^{5/3}+\mathcal{F}\frac{N_{\downarrow}^{2}}{N^{2}}\right],
\end{equation}
where $E_{P}(T)$ and $m^{*}(T)$ are respectively the polaron energy
and effective mass of a \emph{single} impurity at low temperature
$T$. The term proportional to $m/m^{*}$ in the bracket accounts
for the Fermi pressure of the quasi-particle polaron gas, while the
second term with a \emph{defined} Landau parameter $\mathcal{F}$
may be viewed as the interaction energy arising from the effective
polaron-polaron interaction. By taking the partial derivative with
respect to the impurity density, $\partial\Delta E/\partial N_{\downarrow}=\mu_{\downarrow}$,
the impurity chemical potential is given by, 
\begin{equation}
\mu_{\downarrow}=E_{P}\left(T\right)+\frac{m}{m^{*}\left(T\right)}x^{2/3}\varepsilon_{F}+\frac{6}{5}\mathcal{F}\left(T\right)x\varepsilon_{F}+\cdots\label{eq:mudownFL}
\end{equation}
it is important to note that the Landau parameter $\mathcal{F}$ is
always positive and is an effective \emph{repulsive} polaron-polaron
interaction. In Fig. \ref{fig7_FermiLiquid_varyp}, we check the Fermi
liquid description by using the non-self-consistent (upper thick line)
and self-consistent (lower thin line) $T$-matrix theories in the
unitary limit. We fit the calculated impurity chemical potential with
Eq. (\ref{eq:mudownFL}) and then extract the polaron energy $E_{P}$,
inverse effective mass $m/m^{*}$ and Landau parameter $\mathcal{F}$.
The extracted energy and effective mass agree well with these calculated
via Eq. (\ref{eq:Ep}) and Eq. (\ref{eq:Residue}) at the same temperature
$T=0.01T_{F}$. In addition, in the case of the non-self-consistent
$T$-matrix theory, the extracted Landau parameter $\mathcal{F}\simeq0.15$
is close to the value of $\mathcal{F}=0.20$ calculated using the
same ladder approximation \cite{Scazza2017} or the prediction $\mathcal{F}=0.14$
from the fixed-node diffusion Monte Carlo \cite{Pilati2008}.

\begin{figure}[t]
\centering{}\includegraphics[width=0.48\textwidth]{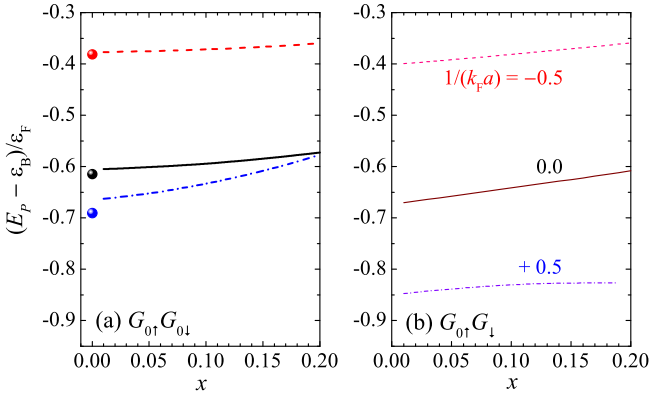}\caption{\label{fig8_ep_varyp_teff001} (color online) The polaron energy as
a function of the impurity concentration $x$ at $T=0.01T_{F}$ and
at three different interaction strengths: $1/(k_{F}a)=-0.5$ (dashed
line), $0$ (solid line), and $0.5$ (dot-dashed line), calculated
by using the non-self-consistent $T$-matrix theory (a, left panel)
and self-consistent $T$-matrix theory (b, right panel). The circles
in (a) show the diagrammatic Monte Carlo result \cite{Vlietinck2013}.}
\end{figure}

\subsection{Density dependence of the polaron energy, mass and residue}

We now consider the density dependence of the quasi-particle properties
of attractive Fermi polarons at zero temperature. It is useful to
note that, the impurity chemical potential is equivalent to the polaron
energy only at zero temperature for a single impurity \cite{Combescot2007}.
In general, the polaron energy defined by Eq. (\ref{eq:Adown}) is
different from the impurity chemical potential. In particular, we
anticipate that the polaron energy is not affected by the many-body
effect of the Fermi pressure term, that is the term $\propto(m/m^{*})x^{2/3}$
in the impurity chemical potential, Eq. (\ref{eq:mudownFL}). However,
the polaron energy may be affected by the residual interaction between
polarons.

Figure~\ref{fig8_ep_varyp_teff001} reports the density dependence
of the polaron energy at three typical interaction strengths and at
$T=0.01T_{F}$, determined by using the two $T$-matrix theories.
As anticipated, the polaron energy does not show the existence of
the Fermi pressure term, which otherwise will lead to a strong density
dependence. The slight increase in the polaron energy with increasing
density, predicted by both theories, could be attributed to the residual
polaron-polaron interaction. According to the non-self-consistent
$T$-matrix theory \cite{Scazza2017}, with increasing interaction
parameter, the Landau parameter $\mathcal{F}$ is small on the BCS
side, takes a maximum at $1/(k_{F}a)\sim0.6$ and finally becomes
small again in the BEC limit. As shown in Fig. \ref{fig8_ep_varyp_teff001}a,
the slopes of the polaron energy as a function of the density $x$
at different interaction strengths are consistent with the interaction
dependence of the Landau parameter $\mathcal{F}$.

\begin{figure}[t]
\centering{}\includegraphics[width=0.48\textwidth]{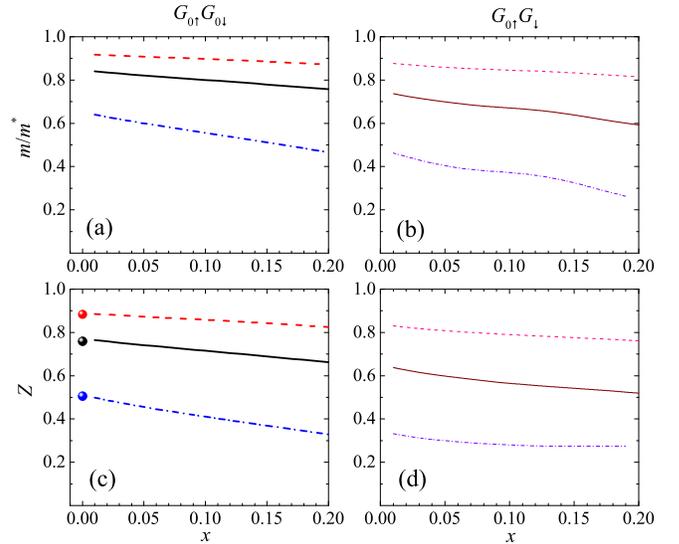}\caption{\label{fig9_MassResidue_varyp_teff001} (color online) The inverse
effective mass (a, b) and the residue (c, d) of attractive Fermi polarons
as a function of the impurity concentration $x$ at $T=0.01T_{F}$
and at different interaction strengths: $1/(k_{F}a)=-0.5$ (dashed
line), $0$ (solid line), and $0.5$ (dot-dashed line), calculated
by using the non-self-consistent $T$-matrix theory (left panel) and
self-consistent $T$-matrix theory (right panel). The circles in (c)
show the diagrammatic Monte Carlo result \cite{Vlietinck2013}.}
\end{figure}

Figure~\ref{fig9_MassResidue_varyp_teff001} presents the density
dependence of the inverse effective mass and residue of Fermi polarons
at $T=0.01T_{F}$. Both quantities decreases with increasing density,
suggesting that the polaronic character is amplified by a finite density.
A reduced polaron residue at nonzero impurity concentration is qualitatively
consistent with the experimental measurement \cite{Schirotzek2009}.
For example, in the unitary limit it was experimentally observed that
$\mathcal{Z}=0.39(9)$ at $5\%$ impurity concentration \cite{Schirotzek2009},
smaller than the variational prediction of $\mathcal{Z}\simeq0.78$
for a single impurity \cite{Chevy2006,Combescot2007}. However, our
results at $x=0.05$, i.e., $\mathcal{Z}\simeq0.74$ from the non-self-consistent
theory and $\mathcal{Z}\simeq0.60$ from the self-consistent theory,
can not quantitatively explain the experimental finding.

\begin{figure}[t]
\begin{centering}
\includegraphics[width=0.48\textwidth]{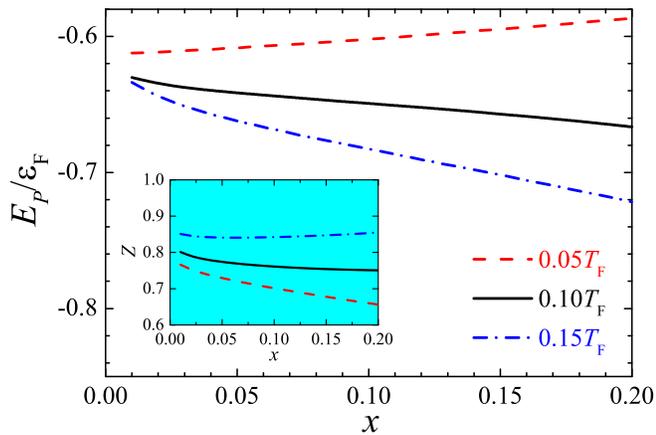} 
\par\end{centering}
\centering{}\caption{\label{fig10_EpResidue_varyp_varyteff} (color online) The polaron
energy as a function of the impurity concentration $x$ in the unitary
limit and at three different temperatures: $T=0.05T_{F}$ (red dashed
line), $0.10T_{F}$ (black solid line) and $0.15T_{F}$ (blue dot-dashed
line). The inset shows the polaron residue.}
\end{figure}

\subsection{Density dependence of the polaron energy and residue at finite temperature}

We now discuss the polaron quasi-particle properties at both nonzero
temperature and nonzero impurity concentration. Figure \ref{fig10_EpResidue_varyp_varyteff}
shows the polaron energy (main figure) and the residue (inset) as
a function of the impurity concentration at three different temperatures:
$T=0.05T_{F}$ (dashed line), $0.10T_{F}$ (solid line), and $0.15T_{F}$
(dot-dashed line). It is interesting that the density dependence qualitatively
changes at higher temperatures. For instance, at $T=0.15T_{F}$ the
polaron energy appears to decrease with increasing density, while
the polaron residue starts to increase. Physically, a reduced polaron
energy at nonzero impurity density indicates an $attractive$ effective
interaction between polarons. Thus, we conclude that the effective
polaron-polaron interaction may change its sign with increasing temperature.

At high temperature, where the polaron becomes more likely an individual,
isolated impurity, the attractive polaronic interaction could be understood
from the induced interaction due to the exchange of medium atoms.
For a weak impurity-medium interaction $U$, it is well known that
such an exchange process leads to an induced interaction \cite{Heiselberg2000,Kinnunen2015}
\begin{equation}
U_{\textrm{ind}}=-U^{2}\chi(\mathbf{q},\omega),
\end{equation}
which should be attractive. Here $\chi(\mathbf{q},\omega)$ is the
density-density response function of the medium atoms with momentum
$\mathbf{q}$ and frequency $\omega$ \cite{Heiselberg2000,Kinnunen2015}.
Our results of a temperature-dependent polaron-polaron interaction
suggest that the weak-coupling picture of induced interactions should
be improved close to zero temperature, in order to accommodate a repulsive
effective interaction between polarons.

\begin{figure}
\begin{centering}
\includegraphics[width=0.48\textwidth]{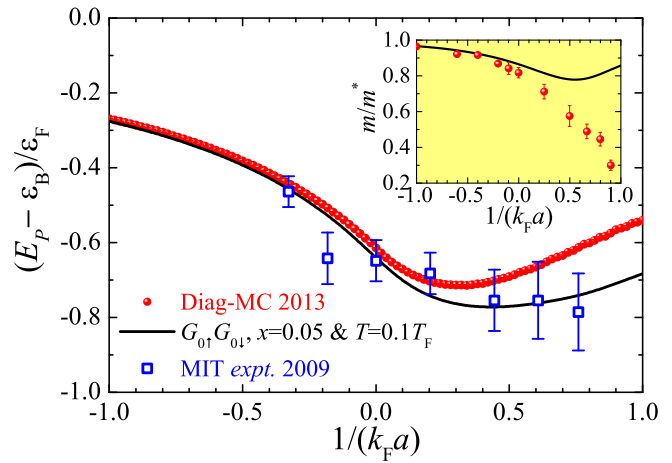} 
\par\end{centering}
\centering{}\caption{\label{fig11_experiments} (color online) The polaron energy at finite
temperature $T=0.10T_{F}$ and at nonzero impurity concentration $x=0.05$,
determined by using the non-self-consistent $T$-matrix theory (black
solid line). For comparison, we show also the diagrammatic Monte Carlo
results at zero temperature (red circles) \cite{Vlietinck2013} and
the experimental data on $^{6}$Li atoms at similar temperature and
impurity concentration (i.e., $T_{\textrm{expt}}=0.14(3)T_{F}$ and
$x_{\textrm{expt}}\simeq0.05$, blue empty squares) \cite{Schirotzek2009}.
The latter is extracted from the review article \cite{Massignan2014}.}
\end{figure}

\subsection{Comparison with Fermi polaron experiments}

An attractive polaron-polaron interaction at nonzero temperature may
lead to a sizable down-shift in the polaron energy at finite impurity
concentration. To show this we report in Fig.~\ref{fig11_experiments}
the polaron energy at $T=0.10T_{F}$ and $x=0.05$, across the whole
the BCS-BEC crossover, calculated by using the non-self-consistent
$T$-matrix theory. The inverse effective mass is shown in the inset.
It is readily seen that the correction due to the combined effect
of temperature and impurity concentration is negligible on the BCS
side and in the unitary limit. However, the correction becomes increasingly
pronounced on the BEC side. The down-shift in the polaron energy is
typically about $0.1\sim0.2\varepsilon_{F}$, and more impressively
the inverse effective mass $m/m^{*}$ becomes less dependent on the
interaction parameter $1/(k_{F}a)$, staying around $m/m^{*}\simeq0.8\sim0.9$.

We now make a comparison with the first attractive Fermi polaron experiment
\cite{Schirotzek2009}. The experiment was carried out at similar
temperature (i.e., $T_{\textrm{expt}}=0.14(3)T_{F}$) and impurity
concentration ($x_{\textrm{expt}}\simeq0.05$). It is encouraging
to see that our theoretical prediction of the polaron energy fits
well with the measured data (empty squares with error bars) \cite{Schirotzek2009},
better than the Diag-MC results for a single impurity at zero-temperature
(solid circles) \cite{Vlietinck2013}, significantly on the BEC side
of the interaction.

Our result of an interaction-insensitive effective mass at finite
temperature is useful to understand the \emph{weak} density dependence
of the rf peak positions for the attractive branch \cite{Schirotzek2009}.
Experimentally, the polaron energies $E_{P\pm}$ for both attractive
and repulsive polarons are measured from the peak position $\Delta_{\pm}$
of the rf-spectroscopy \cite{Scazza2017}, 
\begin{equation}
\Delta_{\pm}=E_{P\pm}-\left(1-\frac{m}{m^{*}}\right)\bar{\varepsilon},
\end{equation}
where the second term on the right-hand-side of the equation reflects
the different dispersion relation of an impurity in the polaron state
and in the third free hyperfine state, and where $\bar{\varepsilon}=\left\langle \hbar^{2}\mathbf{k}^{2}/(2m)\right\rangle $
is the mean kinetic energy per impurity due to the finite impurity
concentration $x\neq0$ (see Eq. (\ref{eq:Adown})). Therefore, if
the effective mass $m^{*}$ of Fermi polarons differs notably from
the bare mass $m$ (i.e., $1-m/m^{*}\gg0$), one can measure $m/m^{*}$
from the dependence of $\Delta_{\pm}$ on $\bar{\varepsilon}$. This
protocol works very well for repulsive Fermi polarons of $^{6}$Li
atoms \cite{Scazza2017}. However, it does not work for attractive
Fermi polarons \cite{Schirotzek2009}, although the variational theory
predicts small enough $m/m^{*}$ for both repulsive and attractive
Fermi polarons at about $1/(k_{F}a)\sim0.6$ at \emph{zero} temperature
\cite{Scazza2017}. As shown in the inset of Fig. \ref{fig11_experiments},
the inverse effective mass of attractive Fermi polarons at finite
temperature actually differs significantly from its zero temperature
value on the experimentally relevant BEC side. The quantity $1-m/m^{*}$
is close to zero and the resulting weak dependence of $\Delta_{-}$
on $\bar{\varepsilon}$ shows experimentally it will be difficult
to extract $m/m^{*}$.

\section{Conclusions}

\label{sec:conc}

In summary, we have presented a systematic investigation of the effects
of finite temperature and finite impurity concentration on the quasi-particle
properties of attractive Fermi polarons. On the BEC side beneath the
Feshbach resonance of the impurity and medium atoms, we have found
that a nonzero temperature, as small as one-tenth of the Fermi temperature,
may lead to a sizable correction to the polaron energy. In this regime
the effective mass of attractive polarons can be reduced significantly,
leading to a weak dependence of the measured resonance peak in the
radio-frequency spectroscopy on the impurity concentration. These
results have been used to better understand the first Fermi polaron
experiment carried out at MIT in 2009 \cite{Schirotzek2009}.

It will be interesting to extend our study to the case of repulsive
Fermi polarons and consider the temperature effect in the recent LENS
experiment with $^{6}$Li atoms \cite{Scazza2017}. To do so, we need
to solve the coupled $T$-matrix equations Eqs. (\ref{eq: gfdown}),
(\ref{eq:selfenergy1}), (\ref{eq:vertexfunction}) and (\ref{Eq:propagator2p})
in the real-frequency domain and determine the single-particle spectral
function of the impurity. Our many-body $T$-matrix theories may also
be extended to address the possible finite temperature effect in Bose
polarons, which have been recently experimentally investigated \cite{Hu2016,Jorgensen2016}. 
\begin{acknowledgments}
We thank very much Kris Van Houcke for kindly sharing their Diag-MC
data in Ref. \cite{Vlietinck2013} with us. HH and XJL acknowledges
the hospitality of Institute for Advanced Study at Tsinghua University,
where a part of the work was done during their visit. Our research
was supported by Australian Research Council's (ARC) Discovery Projects:
DP140100637 and FT140100003 (XJL), FT130100815 and DP170104008 (HH). 
\end{acknowledgments}

\end{document}